\documentclass[english]{article}
\usepackage[T1]{fontenc}
\usepackage[latin9]{inputenc}

\makeatletter
\newcommand{\lyxaddress}[1]{
\par {\raggedright #1
\vspace{1.4em}
\noindent\par}
}

\makeatother

\usepackage{babel}
\begin{document}

\title{\textbf{Interferometric detection of gravitational waves: how can
a wild roam through mindless mathematical laws really be a trek towards
the goal of unification?}}

\author{\textbf{$^{1}$C. Corda, $^{2}$R. Katebi and $^{3}$N. O. Schmidt}}
\maketitle

\lyxaddress{\textbf{$^{1}$}Research Institute for Astronomy and Astrophysics
of Maragha (RIAAM), P.O. Box 55134-441, Maragha, Iran and Dipartimento
di Fisica, Scuola Superiore di Studi Universitari e Ricerca \textquotedbl{}Santa
Rita\textquotedbl{}, Via Severino Delogu, 6 - 00144 Roma Eur, Italy}

\lyxaddress{\textbf{$^{2}$}Department of Physics and Astronomy, Ohio University,
Athens, Ohio 45701, USA }

\lyxaddress{\textbf{$^{3}$}Department of Mathematics, Boise State University,
1910 University Drive, Boise, ID 83725, USA}
\begin{quote}
Essay written for the FQXi Essay Contest 2016: Wandering Towards a
Goal 
\end{quote}
\begin{abstract}
The event GW150914 was the first historical detection of gravitational
waves (GWs). The emergence of this ground-breaking discovery came
not only from incredibly innovative experimental work, but also from
a centennial of theoretical analyses. Many such analyses were performed
by pioneering scientists who had wandered through a wild territory
of mathematical laws. We explore such wandering and explain how it
may impact the grand goal of unification in physics. 
\end{abstract}
In November of 1915 Albert Einstein sent his historical paper on the
general theory of relativity {[}1{]} to the Prussian Academy of Science.
Two subsequent papers by the same Einstein, in 1916 {[}2{]} and in
1918 {[}3{]}, predict that any massive object moving through space-time
will generate GWs. Thereafter, in September of 2015 the Laser Interferometer
Gravitational-Wave Observatory (LIGO) detected the first GW signal
from a binary black hole merger; this remarkable, historical event
is known as GW150914 {[}4{]} and is in alignment with the goal of
establishing a unified field theory of physics. The event GW150914
represented a cornerstone for science and for gravitational physics
in particular. In fact, this remarkable event equipped scientists
with the means to give definitive proof of the existence of GWs, the
existence of black holes having mass greater than 25 solar masses,
and the existence of binary systems of black holes which coalesce
in a time less than the age of the universe {[}4{]}. The one century
period between these two historical events encompasses a great interplay
of experimental and theoretical advancements. Who on Earth would have
guessed that pioneering scientists, who have wandered through a wild
territory of mindless mathematical laws, would ultimately help pave-the-way
to the creation of LIGO and the detection of GW150914? 

In order to take science to the next level, we must first establish
and experimentally-verify a unified field theory of physics so it
can be further applied to disciplines such as chemistry, biology,
engineering, computing, and medicine, etc. Thus, in order to develop
and assess candidates for such a grand unified theory, we must be
able to probe systems of massive objects throughout the universe.
To learn how such systems operate and interact, we must be able to
detect and analyze the GWs that they generate in the first place;
this motivates the hunt for GWs. Detecting GWs is a mighty challenge
because it requires highly-sophisticated technology that is capable
of making extremely precise measurements. Consequently, in order to
advance science and establish a unified field theory, scientists working
in this field need such tools to detect and analyze GWs. Therefore,
the creation of LIGO and the GW150914 detection are colossal, powerful
steps forward in the trek that aims to achieve unification. A great
hope is indeed the future detection of primordial GWs, which could
show that gravity might also be brought into the unification and will
be, in turn, a next great step toward establishing the root of a unified
field theory that may be verified in the laboratory {[}5{]}.

Before we explore aspects of the mindless wandering which ultimately
led to the ground-breaking detection of GW150914 by LIGO, let us briefly
touch on some pertinent background material: what is a GW and which
types of astrophysical objects generate GWs that scientists can detect?
Let us consider an analogy. Suppose that you're standing along the
shore of a vast lake that is flat calm. If you reach down and move
your hand through the water, then your action will cause a disturbance,
or ripple effect, as the generated water waves ensue the trajectory
of your hand and propagate outward in the lake. Albert Einstein's
general theory of relativity predicts a similar effect: if an object
with mass is moving through space-time, then this action will cause
a disturbance as the generated GWs ensue the trajectory of the object
and propagate outward in the universe. Thus, according to the general
theory of relativity, any accelerating object with mass should generate
GWs. But small ripples in space-time would dissipate relatively quickly,
just as small ripples in the lake would fade out before they could
be seen by a second observer standing along the shore at a sufficiently
far distance. Therefore, in order to detect GWs on Earth that are
generated from distant sources throughout the universe, scientists
search for enormously massive objects, such as neutron stars or black
holes, that are capable of generating GWs that propagate all the way
to Earth. 

In 1974 Russel A. Hulse and Joseph H. Taylor discovered the Hulse-Taylor
binary (or PSR 1913+16) using the Arecibo radio telescope {[}6{]}.
The Hulse-Taylor binary is a compact star system consisting of two
neutron stars (one of which is a pulsar - i.e. a neutron star emitting
electromagnetic radiation) that orbit around a common center of mass.
The Hulse-Taylor binary was the first binary star system to be observed
and is regarded as the first indirect proof of the existence of GWs
as predicted by the general theory of relativity. Even though the
first efforts at direct GW detection started long before the Hulse-Taylor
binary (see the recent paper by the Nobel Laureate G. F. Smoot and
collaborators on the history of GW research {[}7{]}), this astronomical
revelation generated intense excitement in the physics community while
further motivating efforts to create new technology for direct GW
detection. In fact, this historical finding was so significant that
it earned Hulse and Taylor the 1993 Nobel Prize in Physics {[}8{]}. 

On one hand, the event GW150914 and the subsequent event GW151226,
which is the second direct detection of GWs from a 22 ­solar-­mass
binary black hole coalescence {[}9{]} are considered to be one of
the greatest triumphs of experimental physics because they represent
the most precise experimental measurements in the whole history of
science. In fact, they involve measuring distances on the order of
$10^{-18}$ meters. This is a distance shorter than the proton's radius!
In a recent interview during a trip in Italy, the famous theoretical
physicist Kip Thorne, who is one of the LIGO's Founding Fathers and
is, in turn, a candidate for the Nobel Prize in Physics for the direct
detection of GWs, claimed, with a very remarkable modesty that {[}10{]}
(translated from the Italian language) ``\emph{The next Nobel Prize
for Physics? It must be assigned to the gravitational waves, but I
do not deserve it. The real heroes of this event (the detection of
gravitational waves) are the experimental physicists who resolved
all the practical problems of a very complex experiment making possible
this discovery. I am not among them.''.}

On the other hand, the goal of this Essay is to stress that the event
GW150914 and the subsequent event GW151226 arise not only from extremely
precise experimental work, as it has been emphasized by Thorne, but
also from a centennial of theoretical analyses which have been performed
through lots of mindless mathematical laws. 

Einstein predicted the existence of GWs in his theoretical, historical
paper {[}2{]} and improved his analysis two years later {[}3{]}, claiming
that this second paper permitted him to correct a trivial mistake
in his previous work {[}2{]}. In any case, Einstein's position on
the existence or non-existence of GWs changed various times. In 1936
Einstein wrote to Max Born {[}11{]}: ``\emph{Together with a young
collaborator (Nathan Rosen), I arrived at the interesting result that
gravitational waves do not exist, though they had been assumed a certainty
to the first approximation. This shows that the non-linear general
relativistic field equations can tell us more or, rather, limit us
more than we have believed up to now}''. Einstein submitted this
research titled \textquotedblleft \emph{Do Gravitational Waves Exist?}\textquotedblright{}
to the Physical Review with Rosen as the co-author {[}12{]}. Although
the original version of this manuscript no longer exists, one can
infer from the letter of Einstein to Born {[}11, 12{]} that Einstein
and Rosen answered \textquotedbl{}No\textquotedbl{} in response to
the question of the title. Despite Einstein's great eminence and fame,
the Physical Review returned the paper to Einstein with a critical
review and the kind request that the journal's Editor, who was the
physicist John Torrence Tate Sr., \textquotedblleft \emph{would be
glad to have {[}Einstein\textquoteright s{]} reaction to the various
comments and criticisms the referee has made}\textquotedblright{}
{[}12{]}. When Einstein received the returned paper with the critical
review he became infuriated and decided to ultimately withdraw the
paper from the Physical Review with a very irritated letter (Einstein
was so furious that he wrote the letter in Germany instead of in English!).
Details of this curious story can be found in {[}12{]}. In any case,
Einstein, who decided to publish the paper with the Journal of the
Franklin Institute in Philadelphia {[}13{]}, again reversed his opinion
on the existence of GWs in 1937. In fact, Einstein's new collaborator,
Infeld, discovered a mistake in the paper of Einstein and Rosen during
a discussion with the relativist Howard Percy Robertson {[}12, 14{]},
who is famous for his works in cosmology (he was co-author of the
widely known Robertson-Walker metric). Infeld reported his discussion
with Robertson to Einstein. This time Einstein not only agreed with
the arguments of Infeld and Robertson, but also added that he had
coincidentally and independently found another mistake in the paper
that he wrote together with Rosen {[}12, 14{]}. At that time the paper
was in the phase of proofs for the Journal of the Franklin Institute.
Thus, despite that the journal had already accepted the paper in its
original form, Einstein was forced to explain that fundamental changes
in the paper were required because the consequences of the equations
derived in the manuscript were incorrect {[}12{]}. Then, the paper
of Einstein and Rosen was published with radically altered conclusions
{[}13{]}. Differently from Einstein, Rosen did not change his opinion
on the non-existence of GWs {[}12{]}. In fact, he was not happy with
the paper {[}13{]} and published a paper in a Soviet journal by claiming
the non-existence of GWs {[}15{]}. In the following year, Einstein
reversed his opinion one more time. In fact, in 1938 Einstein wrote
a paper together with Infeld and the mathematician and physicist Banesh
Hoffmann {[}16{]}. This was a (fruitless) attempt to find a theory
which could unify gravity and electromagnetism, where one of the assumptions
of the paper was that GWs should not exist {[}16{]}. Thus, in general,
Einstein's attitude on the existence or non-existence of GWs was of
substantial uncertainty. This was stressed by the same Einstein in
a lecture that he delivered to the University of Princeton exactly
one day after he corrected his paper {[}13{]}. In fact, he concluded
the lecture by saying ``\emph{If you ask me whether there are gravitational
waves or not, I must answer that I do not know. But it is a highly
interesting problem}'' {[}12, 14{]}. 

A key event in GW research occurred in the 1950s, when the famous
astrophysicist Hermann Bondi, with his collaborators Felix Arnold
Edward Pirani and Ivor Robinson, published the fundamental paper {[}17{]}.
In that work, they showed that Rosen's arguments in {[}15{]}, which
claimed that GWs do not exist, were incorrect. Furthermore, they also
correctly predicted the effect that would eventually be used in the
future by LIGO to detect the real GWs of GW150914 and GW151226. The
most remarkable contribution on this issue was by Pirani, who also
wrote the important papers {[}18 - 20{]}. 

Thus, for the theme of this Essay, we note that for more than 40 years
- i.e. from Einstein's prevision in {[}2,3{]} to the work of Pirani
and Bondi {[}17 - 20{]} - the primary goal of GW detection was not
yet well-defined. Instead, we've had a gigantic and intriguing, but
also wild and controversial, debate regarding the existence or non-existence
of GWs through a hefty amount of theoretical work over a vast territory
of mindless mathematical laws. In this case, identifying the real
goal to be approached (the detection of GWs) was a greatly disputed
issue. In order to ultimately realize the significance of GW detection
and develop the mathematical formalism to describe the GW phenomena,
numerous pioneering scientists invested an enormous amount of time
and energy to wander through this mindless territory by asking questions,
considering hypotheses and conducting thought experiments. We indeed
cited only the most important contributions to the debate on GWs,
which also involved the work of Eddington {[}21{]}, Beck {[}22{]},
Baldwin and Jeffery {[}23{]}. Despite Einstein's claim that, even
admitting the concrete existence of GWs, their detection will always
be impossible because of the very weak coupling between matter and
GWs {[}3{]}, Pirani and collaborators {[}17 - 20{]} predicted a GW
effect that could be observed. They indeed proposed the geodesic deviation
equation as a tool for designing a practical GW detector. In other
words, if a GW propagates in a region of space-time where two free-falling
test masses are present, the GW effect will drive the masses to oscillate. 

Recently, one of us, C. Corda, generalized the work in {[}17 - 20{]}
to the framework of extended theories of gravity {[}24 - 26{]}, also
together with collaborators {[}27{]}. In fact, the motion of the test
masses due to a GW in extended theories of gravity is different with
respect to the motion of test masses due to a GW in the general theory
of relativity {[}24 - 27{]}. This is because in the standard general
theory of relativity one finds only two different, independent GW
polarizations, while in extended theories of gravity the independent
GW polarizations are at least three {[}24 - 27{]}. The results in
{[}24 - 27{]} could become very important in the framework of the
nascent GW astronomy in order to ultimately discriminate between the
general theory of relativity and extended theories of gravity, with
important consequences concerning the final unification of theories.
An extension of the general theory of relativity could indeed be necessary
in order to achieve such a prestigious goal - see {[}24{]} for details.

The greatest problem in GW interferometric detectors is that the ``signal'',
which is the motion of the test masses, is very weak. In fact, let
us consider a GW which originates from an astrophysical source and
propagates in a region of space-time where two test masses stay separated
by a distance on the order of a few kilometers. The GW drives such
test masses to oscillate with an oscillation amplitude of order of
$10^{-18}$ meters. In order to achieve this extremely difficult measure,
GW physicists use the so-called \emph{intereferometers.} These are
extremely precise ``yardsticks'' that use the properties of light
in order to realize this almost impossible measure. Thus, on one hand,
the main task for experimental physicists is to figure out how to
reduce the potential noise interference that makes such tiny measurements
so difficult to perform in practice. The most important source of
noise for interferometric GW detectors is the seismic noise, but many
additional sources of noise are present; see {[}28{]} for details.
On the other hand, reducing the potential impact of various sources
of noise has not been sufficient in order to guarantee the GW detection.
We have seen that an enormous amount of computations and mathematical
laws were necessary only to identify the goal of GW detection. During
the development of interferometric GW detectors, which starts from
the original intuitions of the Soviet physicists Mikhail Evgen'evich
Gertsenshtein and Vladislav Ivanovich in the early 1960s {[}29{]}
and continues until the LIGO discovery {[}4{]}, an immense number
of simulations and data analyses have been indeed performed by using
various mindless mathematical laws. An important and very useful tool
has indeed been numerical relativity, which is the branch of the general
theory of relativity that creates algorithms and uses numerical methods
to analyze and potentially solve problems. To achieve the detection
of the event GW150914, many researchers have worked hard to obtain
numerical solutions to the problem of a binary black hole system,
which enables them to get increasingly accurate computational results
that describe the GWs emitted by such a system {[}4{]}. In fact, scientists
first generated the wave forms by simulating binary black holes, black
hole - neutron stars and neutron star - neutron star mergers by employing
numerical relativity and powerful super computers in order to have
many template waveforms for checking the possible detections {[}30
- 32{]}. Thus, the work of experimental physicists must be complementary
to the work of theorists.

Concerning the previously cited possibility of ultimately discriminating
between the general theory of relativity and extended theories of
gravity, only a perfect knowledge of the motion of the test masses,
which are the beamsplitter and the mirrors of the interferometer,
will permit one to determine if the general theory of relativity is
the definitive theory of gravity. At the present time, the sensitivity
of the current ground-based GW interferometers is not high enough
to determine the motion of the test masses with an absolute precision.
A network including interferometers with different orientations is
indeed required and we're hoping that future advancements in ground-based
projects and space-based projects will have a sufficiently high sensitivity.
Such advancements would enable physicists to determine, with absolute
precision, the direction of GW propagation and the motion of the various
involved mirrors. In other words, in the nascent GW astronomy we hope
not only to obtain new, precise, astrophysical information, but we
also hope to be able to obtain a perfect knowledge of the motion of
the test masses. Such advances in GW technology would equip us with
the means and results to ultimately confirm the general theory of
relativity or, alternatively, to ultimately clarify that the general
theory of relativity must be extended. This ambitious result, we observe,
will only be obtained through a correct mixture of technological innovation,
collaboration, debate, and wild roams through mindless mathematical
laws with adherence to the scientific method. This achievement will
surely be a great step forward in the trek towards the grand unification
of physics.

\section*{Acknowledgements }

Christian Corda has been supported financially by the Research Institute
for Astronomy and Astrophysics of Maragha (RIAAM), Iran.

\section*{References}

{[}1{]} A. Einstein, Sitzungsber. Preuss. Akad. Wiss. Berlin (Math.
Phys.) 778 (1915). Addendum: Sitzungsber. Preuss. Akad. Wiss. Berlin
(Math. Phys.) 799 (1915).

{[}2{]} A. Einstein, Sitzungsber. K. Preuss. Akad. Wiss. \textbf{1},
688 (1916).

{[}3{]} A. Einstein, Sitzungsber. K. Preuss. Akad. Wiss. \textbf{1},
154 (1918).

{[}4{]} B. Abbott et al. (LIGO Scientific Collaboration and Virgo
Collaboration), Phys. Rev. Lett. \textbf{116}, 061102 (2016).

{[}5{]} L. M. Krauss and F. Wilczek, Int. J. Mod. Phys. D \textbf{23},
1441001 (2014).

{[}6{]} R. A. Hulse and J. H. Taylor, Astrophys. J. \textbf{195},
L51 (1975).

{[}7{]} J. L. Cervantes-­Cota, S. Galindo-­Uribarri e G. F. Smoot,
Universe, 2(3), 22 (2016).

{[}8{]} http://www.nobelprize.org/nobel\_prizes/physics/laureates/1993/press.html.

{[}9{]} B. P. Abbott et al. (LIGO Scientific Collaboration and Virgo
Collaboration), Phys. Rev. Lett. \textbf{116}, 241103 (2016).

{[}10{]} L. Fraioli, http://ricerca.repubblica.it/repubblica/archivio/repubblica/2016/06/26/non-sono-un-eroe-in-corsa-per-il-nobel-ma-so-che-luniverso-segreti19.html.

{[}11{]} A. Einstein, The Born\textendash Einstein Letters: \emph{Friendship,
Politics, and Physics in Uncertain Times}, p. 122, MacMillan, New
York (2005). 

{[}12{]} D. Kennefick, Phys. Tod., \textbf{58 (9)}, 43 (2005)

{[}13{]} A. Einstein, N. Rosen, J. Franklin Inst. \textbf{223}, 43
(1937).

{[}14{]} L. Infeld, \emph{Quest: An Autobiography}, Chelsea, New York
(1980).

{[}15{]} N. Rosen, Phys. Z. Sowjetunion 12, 366 (1937).

{[}16{]} A. Einstein, L. Infeld and B. Hoffmann, Ann. Math., Sec.
Ser., \textbf{39 (1)}, 65 (1938).

{[}17{]} H. Bondi, F. A. E. Pirani e I. Robinson, Proc. Roy. Soc.,
A \textbf{251}, 519 (1959).

{[}18{]}F.A.E. Pirani, Third Award at the Gravity Research Foundation
Competition 1957, www.gravityresearchfoundation.org/pdf/awarded/1957/pirani.pdf
(1957).

{[}19{]} F. A. E. Pirani, Acta Physica Polonica 15, 389 (1956).

{[}20{]} F. A. E. Pirani, Phys. Rev. 105, 1089 (1957).

{[}21{]} A. Eddington, Proc. Roy. Soc. A, \textbf{102}, 716 (1922).

{[}22{]} G. Beck, Z. Phys. \textbf{33}, 713 (1925).

{[}23{]} O. R. Baldwin, G. B. Jeffery, Proc. Phys. Soc. London, Sect.
A 111, 95 (1926).

{[}24{]} C. Corda, Int. J. Mod. Phys. D, \textbf{18}, 2275 (2009).

{[}25{]} C. Corda, Phys. Rev. D \textbf{83}, 062002 (2011).

{[}26{]} C. Corda, JCAP, \textbf{0704}, 009 (2007). 

{[}27{]} S. Capozziello, C. Corda and M. F. De Laurentis, Phys. Lett.
B \textbf{669}, 255 (2008). 

{[}28{]} P. R Saulson,\emph{ Fundamentals of Interferometric Gravitational
Wave Detectors}, World Scientific (1994).

{[}29{]} The LIGO Collaboration, \emph{A Brief History of LIGO}, 

https://www.ligo.caltech.edu/system/media\_files/binaries/313/original/LIGOHistory.pdf

{[}30{]} B. P. Abbott et al., Phys. Rev. X \textbf{6}, 041014 (2016). 

{[}31{]} A. H. Mroué et al., Phys. Rev. Lett. \textbf{111}, 241104
(2013).

{[}32{]} G. Lovelace et al., Class. Quantum Grav. \textbf{32}, 065007
(2015).
\end{document}